%% file: main.tex
\documentclass[sigconf, nonacm]{acmart}
\pdfoutput=1
\newcommand\vldbdoi{XX.XX/XXX.XX}
\newcommand\vldbpages{XXX-XXX}
\newcommand\vldbvolume{}
\newcommand\vldbissue{}
\newcommand\vldbyear{}
\newcommand\vldbauthors{\authors}
\newcommand\vldbtitle{\shorttitle} 
\newcommand\vldbavailabilityurl{https://github.com/madelonhulsebos/AdaTyper}
\newcommand\vldbpagestyle{plain}
\newcommand{\ignore}[1]{}
\usepackage{graphicx}
\usepackage[utf8]{inputenc}
\usepackage{times}
\usepackage{listings}
\usepackage{courier}
\usepackage{svg}
\usepackage{float}
\usepackage{subcaption}
\usepackage{balance}
\usepackage{amsmath}
\usepackage{url}
\usepackage[export]{adjustbox}
\usepackage{xspace}
\usepackage{pifont}
\usepackage{outlines}
\usepackage{adjustbox}
\usepackage{multirow}
\usepackage{comment}
\usepackage{enumitem}
\usepackage{todonotes}
\usepackage{fancybox}
\usepackage[boxed,linesnumbered]{algorithm2e}
\usepackage{tablefootnote}

\renewcommand{\paragraph}[1]{\vspace{0.2\baselineskip}\noindent\textbf{#1.}\hspace{0.1cm}}

\newcommand{\system}{{\sc AdaTyper}\xspace}
\newcommand{\type}[1]{\texttt{#1}}

\renewcommand\footnotetextcopyrightpermission[1]{}

\SetAlCapSkip{0.5em}
\SetAlCapHSkip{-0.5cm}

\begin{document}

\title{AdaTyper: Adaptive Semantic Column Type Detection [Scalable Data Science]}

\author{Madelon Hulsebos}
\authornote{Work started and partially completed at Sigma Computing.}
\affiliation{%
  \institution{University of Amsterdam}
  \city{Amsterdam}
  \country{Netherlands}
}
\email{m.hulsebos@uva.nl}

\author{Paul Groth}
\affiliation{
  \institution{University of Amsterdam}
  \city{Amsterdam}
  \country{Netherlands}
}
\email{p.t.groth@uva.nl}

\author{\c{C}a\u{g}atay Demiralp}
\authornotemark[1]
\affiliation{%
  \institution{MIT CSAIL}
  \city{Cambridge}
  \country{USA}
}
\email{cagatay@csail.mit.edu}

\input{00_abstract.tex}

\maketitle

\pagestyle{\vldbpagestyle}
\begingroup\small\noindent\raggedright\textbf{PVLDB Reference Format:}\\
\vldbauthors. \vldbtitle. PVLDB, \vldbvolume(\vldbissue): \vldbpages, \vldbyear. 
\href{https://doi.org/\vldbdoi}{doi:\vldbdoi}
\endgroup
\begingroup
\renewcommand\thefootnote{}\footnote{\noindent
This work is licensed under the Creative Commons BY-NC-ND 4.0 International License. Visit \url{https://creativecommons.org/licenses/by-nc-nd/4.0/} to view a copy of this license. For any use beyond those covered by this license, obtain permission by emailing \href{mailto:info@vldb.org}{info@vldb.org}. Copyright is held by the owner/author(s). Publication rights licensed to the VLDB Endowment. \\
\raggedright Proceedings of the VLDB Endowment, Vol. \vldbvolume, No. \vldbissue\ %
ISSN 2150-8097. \\
\href{https://doi.org/\vldbdoi}{doi:\vldbdoi} \\
}\addtocounter{footnote}{-1}\endgroup

\ifdefempty{\vldbavailabilityurl}{}{
\vspace{.3cm}
\begingroup\small\noindent\raggedright\textbf{PVLDB Artifact Availability:}\\
The source code, data, and/or other artifacts have been made available at \url{\vldbavailabilityurl}.
\endgroup
}



\input{01_introduction}
\input{02_notation}

\input{03_adaptation}
\input{05_adatyper}

\input{06_evaluation}
\input{07_related_work}

\input{08_conclusion}
\input{acknowledgements}

\bibliographystyle{ACM-Reference-Format}
\bibliography{main}


\balance

\end{document}

%% file: 00_abstract.tex
\begin{abstract}
Understanding the semantics of relational tables is instrumental for automation in data exploration and preparation systems. A key source for understanding a table is the semantics of its columns. 
With the rise of deep learning, learned table representations are now available, which can be applied for semantic type detection and achieve good performance on benchmarks. Nevertheless, we observe a gap between this performance and its applicability in practice. In this paper, we propose \system to address one of the most critical deployment challenges: adaptation. \system uses weak-supervision to adapt a hybrid type predictor towards new semantic types and shifted data distributions, at inference time, using minimal human feedback. The hybrid type predictor of \system, combines rule-based methods and a light machine learning model for semantic column type detection. We evaluate the adaptation performance of \system on real-world database tables hand-annotated with semantic column types through crowdsourcing, and find that the f1-score improves for new and existing types. \system approaches an average precision of 0.6 after only seeing 5 examples, significantly outperforming existing adaptation methods based on human-provided regular expressions or dictionaries.
\end{abstract}

%% file: 01_introduction.tex
\section{Introduction}

\begin{figure}
    \centering
    \includegraphics[width=0.86\columnwidth]{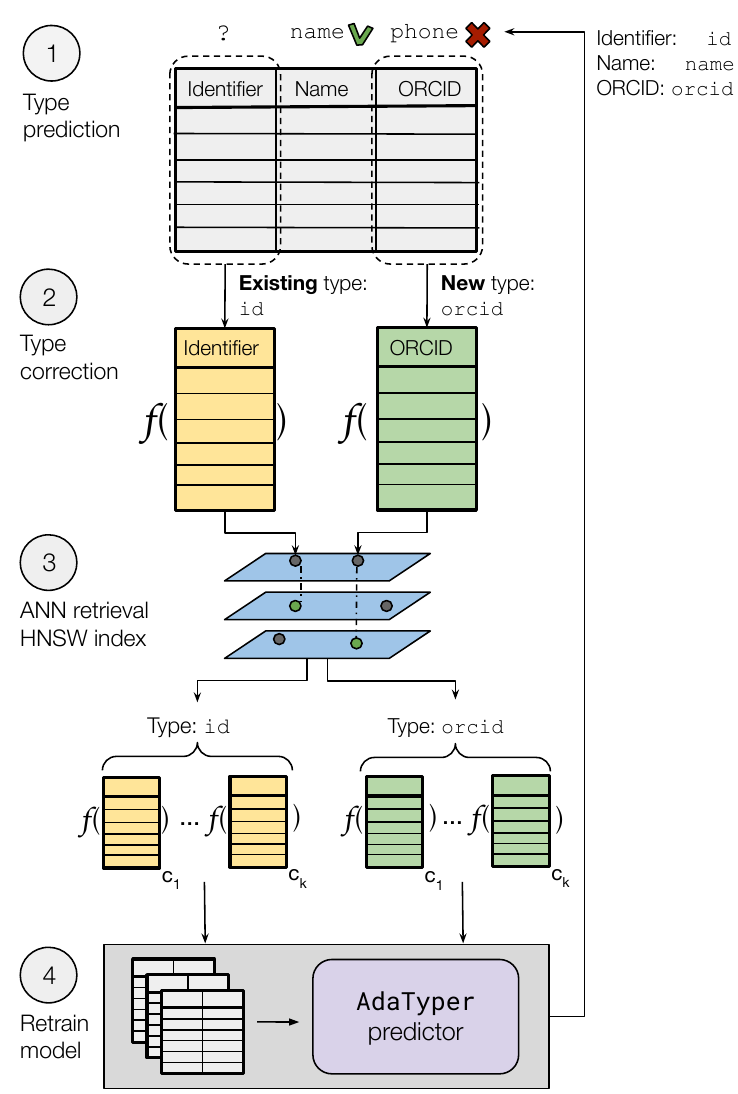}
    \vspace{-0.3cm}
    \caption{\small{Overview of \system for adaptive semantic column type detection where \ding{172} the type predictor yields initial column types \type{name} and \type{phone}, and none for the first column. \ding{173} Upon incorrect or missing type predictions, a human corrects the column type with a type from the catalog (e.g. \type{ID}) or defines a new type (e.g. \type{ORCID}). \ding{174} T5 embeddings of the associated columns, \textit{f}(.), are used to retrieve $k$ approximate nearest neighboring (ANN) columns $c_1,...,c_k$ for each type. \ding{175} The columns are then added to the training corpus as instances of the types \type{ID} and \type{ORCID}, after which the type predictor is retrained.}}
    \label{fig:adatyper-overview}
    \vspace{-0.7cm}
\end{figure}


The data used and generated by enterprise applications are processed and stored overwhelmingly as relational tables. Table understanding aims to surface the semantics of tables to improve tasks ranging from data integration to data visualization. Table understanding~\cite{tableunderstandingtutorial2021} in this context refers to tasks such as table topic inference, entity resolution, and semantic column type detection, which is the focus of this work. We define the task of semantic column type detection in tables as:

\paragraph{Semantic column type detection} \textit{Given a table $x$ with columns $\{c_1, c_2, \dots, c_N\}$, where $N$ is the number of columns in $x$. The objective is to obtain a predictor $P$ to predict accurate semantic types $\{t_1, t_2, \dots, t_N\}$, from a type catalog $\mathcal{T}$, such that $t_1$ accurately reflects the semantics of $c_1$, $t_2$ reflects the semantics of $c_2$, etc.}\vspace{0.15cm}

Analogous to types in programming languages, the semantic type of a column dictates the operations that can be performed on it at runtime and provides useful table metadata. Semantic column types have been instrumental in supporting column-level data preparation and exploration tasks. For automating data preparation~\cite{chu2015katara}, for example, semantic types enable capturing data errors and normalizing values (e.g. capitalizing values that represent names). Similarly, semantic types can be integrated into data visualization recommendation systems, e.g., to inform suitable visualizations for a set of table columns~\cite{hu2019vizml}. Moreover, semantic column types provide table metadata which, beyond data cataloging, helps improve data exploration tasks such as data search~\cite{webtablestenyears}.

Recent work has introduced deep learning to table understanding tasks like semantic type detection~\cite{hulsebos2019sherlock, zhang2020sato}. Using transformers and language models, researchers have also proposed pretrained table models such as TURL~\cite{turl}, TCN~\cite{wang2021tcn}, and TABBIE~\cite{iida2021tabbie}, obtaining excellent accuracy on this task. While there is a heightened research interest in table understanding models, little is known about their applicability in practice due to their limited deployment.

Instead, most commercial data systems primarily rely on methods such as regular expression matching or dictionary lookups for detecting a limited set of semantic types~\cite{talend:type-system, trifacta:type-system, googledatastudio:semantic-type}.
Some systems, e.g. Tamr~\cite{stonebraker2013tamr}, made the step towards machine learning but use user-specific and low-capacity models due to, for example, a lack of training data~\cite{stonebraker2018dataintegration}. Motivated by the potential of practical applications of pretrained table understanding models, we focus our attention on making these models effective in practice, using semantic type detection task as a canonical task.




Feedback from organizations on the early ML-based type detection method Sherlock~\cite{hulsebos2019sherlock}, highlighted the need for adapting type detection systems towards different sets of tables and semantic types specific to domains. Existing data preparation and visualization systems (e.g. Trifacta and Google Data Studio) enable custom types but require manual configuration of regular expressions or dictionaries~\cite{trifacta:type-system, googledatastudio:semantic-type}.
Moreover, adapting learned semantic type detection models  requires new labeled datasets that represent the new types well, which often requires complex labeling procedures and long retraining procedures.
Although the language models tailored for tabular tasks~\cite{turl,yin2020tabert} are meant to relieve these burdens, we observe that these models are not straightforward to fine-tune and still require a significant set of labeled data.


In response, we introduce \system, an adaptive semantic column type detection system. \system incorporates a type predictor trained on GitTables~\cite{hulsebos2023gittables}, providing representative tables and relevant semantic types. This predictor combines pragmatic value and 
pattern matching estimators with a learned model to establish high precision and broad coverage of semantic types. Each individual estimator generates confidence scores to filter out predictions that don't conform to an acceptable error rate. Besides 10 generally relevant semantic types that we use to train the model, we include a ``background type'' to make the model recognize out-of-distribution data. Each estimator of the \system adapts iteratively to a user’s context through weak-supervision requiring minimal user effort (Figure~\ref{fig:adatyper-overview}). That is, \system takes light-weight user feedback to generate custom training data for iterative model training.

We evaluate the adaptation performance of \system on real-world database tables hand-annotated with semantic column types through crowdsourcing, and find that the f1-score improves for new and existing types. After seeing only 5 examples, \system approaches an average precision of 0.6, significantly outperforming existing adaptation methods based on human-provided regular expressions or dictionaries.

In summary, our contributions are as follows:
\begin{enumerate}[leftmargin=0.5cm]

    \item \system: a semantic type predictor combining a learned estimator with pragmatic heuristics, which all adapt towards new semantic types as well as shifted data distributions.
    The \system Predictor is trained on 10 semantic types across 15K tables from GitTables. We publish the code on GitHub\footnote{\url{https://github.com/madelonhulsebos/AdaTyper}}.
    
    \item A human-annotated version of the 1K tables of the CTU Prague Relational Learning Repository~\cite{motl2015ctu}, with 26 general semantic types. We open-source the code of the web application for retrieving table column annotations from humans.
    
    \item An in-depth analysis of the \system Adapter adaptation performance towards new semantic types and out-of-distribution data using the human-annotated CTU dataset.
\end{enumerate}

%% file: 02_notation.tex
\section{Notation}

Throughout the paper we adhere to the following notation:

\begin{table}[h!]
    \centering
    \vspace{-0.1cm}
    \begin{tabular}{l  l }
        \toprule
        $x$ & table\\
        $c_i$ & $i^{th}$ column in $x$\\
        $N$ & number of columns in $x$\\
        $t$ & semantic type\\
        $\mathcal{T}$ & semantic type catalog\\
        $Z$ & the number of semantic types in $\mathcal{T}$, i.e. $|\mathcal{T}|$\\
        $P$ & semantic type predictor\\
        $\mathcal{X}$ & set of tables labeled with semantic types\\
        $c_e$ & embedding of column $c$\\
        $\mathcal{C}_e$ & set of column embeddings\\
        $\mathcal{H}$ & Hierarchical Navigable Small World index\\
        $a_c$ & a semantic type annotation for column $c$ \\
        \bottomrule
    \end{tabular}
    \label{tab:notation}
    \vspace{-0.2cm}
\end{table}

%% file: 03_adaptation.tex
\section{The Challenge of Adaptation}\label{sec:challenges-practice}

\begin{figure}
\includegraphics[width=\columnwidth]{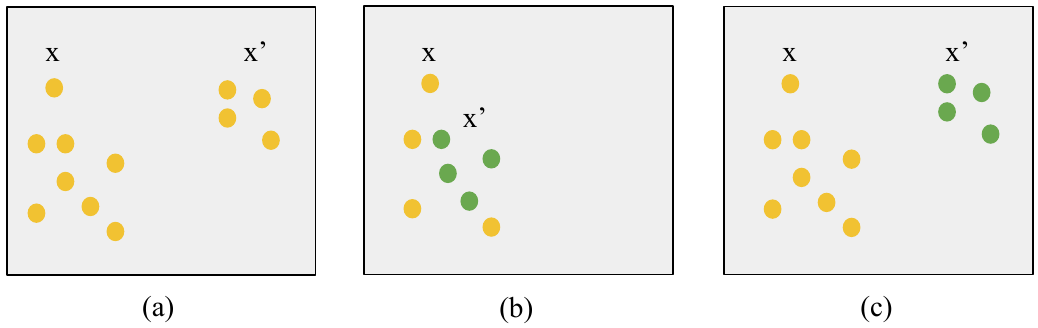}
\vspace{-0.5cm}
\caption{We consider three variants of data-shift: (a) covariate-shift when the data distribution of class $x$ changes on inference time to $x'$, (b) label-shift when the class of $x$ (yellow) at inference time overlaps with samples $x'$ of a new class (green), (c) out-of-distribution data when samples $x'$ of an unseen class are encountered that a model was not trained on. Circles represent table columns, and colors indicate their semantic types (classes), where \textit{yellow} reflects a known semantic type and \textit{green} an unknown semantic type. Columns $x$ are from the training data distribution, and $x'$ from the data distribution at inference time.}
\label{fig:adaptation-scenarios}
\end{figure}

One of the key challenges for deploying learned table models is that model predictions should adapt to the data distributions and semantics in the domain of deployment. 
This problem is particularly apparent in relational data, given that the same values may be tied to the different domain-specific semantic context of users. This is in contrast with adjacent fields, e.g. computer vision, where, for example, an image of a cat represents, invariably, a cat. Hence, in order to provide the requisite functionality, a semantic column type detection system is expected to adapt to a user's context and improve over time. If the system fails to do so, it might lose relevance to a user, leading to a potential decrease in satisfaction or engagement. We formulate this problem as follows:

\paragraph{Adaptive semantic column type detection} \textit{Given an unseen table $x_u$ with columns $\{c_1, c_2, \dots, c_N\}$ and initial semantic type predictions generated by type predictor $P$ per table column $\{t_1, t_2, \dots, t_Z\} \ \in$ type catalog $\mathcal{T}$ of size $Z$. Consider $c_2$ at time $i$ being incorrectly labeled with semantic type $t_1$ or no type at all. The user corrects or provides the semantic type for $c_2$, and chooses an existing type from $\mathcal{T}$ or adds a new type $t_{Z+1}$ to $\mathcal{T}$. The goal of adaptation is to correct type predictor $P$ so that, at time $i+1$, $c_2$ is correctly predicted to be of the provided semantic type from $\mathcal{T}$.\\}

Specifically, there are three cases of data shift that can occur between the training data and the data in the user context that require adaptation. First, the distribution of table values present in the training data may differ from that of the tables at inference-time, which is referred to as covariate-shift~\cite{kouw2018transferlearning} (Fig.~ \ref{fig:adaptation-scenarios}a). For example, in the case of semantic column type detection, a column with the name ``ID'' might contain values not previously seen for the semantic type \type{ID}.
Furthermore, values associated with a certain semantic type in the training dataset could correspond to a different type within the user's context, also called label-shift~\cite{rabanser2018failing}(Fig.~\ref{fig:adaptation-scenarios}b). For example, a column with predicted semantic type \type{ID} might actually correspond to the \type{phone number} type within the user's context. Finally, we may encounter out-of-distribution data (Fig.~\ref{fig:adaptation-scenarios}c). A user might have tables with semantic types far from the training distribution~\cite{rabanser2018failing}. 

Early information retrieval systems have shown that iteratively learning from interactive feedback is an efficient way to accomplish contextualization~\cite{tamine2010evaluation, joachims2007search}. Such interactions should be easy and require minimal time and input, to maximize the effectiveness of the feedback cycle. Existing systems often, however, require complicated manual configurations that assume deeper knowledge of the expected data values or technical know-how, e.g. for specifying regular expressions~\cite{talend:type-system, googledatastudio:semantic-type}.

Moreover, prior benchmarks of deep learning models~\cite{hulsebos2019sherlock} for semantic type detection identified that regular expressions inferred from data do not generalize well to a variety of semantic types. As a result, adapting type predictors based on manual or automatically inferred regular expressions can yield low precision leading to many errors. These systems also do not estimate posterior probability densities, making it difficult to infer confidence scores. We need more adequate methods to address these problems.

%% file: 05_adatyper.tex
\section{AdaTyper}\label{sec:sigmatyper} 
\system is a system that implements adaptive semantic column type detection using weak-supervision with minimal  feedback from humans. In this section, we describe the training data and base type predictor (\system Predictor), and describe the design of the component for domain adaptation (\system Adapter).

\subsection{Training data: tables and types}\label{sec:training-data}
As traditional large-scale table corpora, such as WebTables~\cite{webtables2012} and WikiTables~\cite{bhagavatula2013wikitables} do not extend well to database tables~\cite{langenecker2021towards, hulsebos2023gittables}, newer data sources~\cite{herzig2021open, hulsebos2023gittables, vogelsgesang2018getreal} aim at representing database-like tables. Since \system is intended to operate on tables found in varied organizations, we use tables from GitTables~\cite{hulsebos2023gittables} as our training dataset. GitTables has been introduced to address the need for database tables to train learned models over tables for broader applicability, including in enterprises.
 
Given our focus on adaptation to organization data contexts, we take an initial type catalog $\mathcal{T}$ with 10 types covering geographic, demographic, and business types. Tables in GitTables are annotated with over 500 semantic types from DBpedia~\cite{dbpedia} and Schema.org~\cite{guha2016schema}. We select the semantic types from the DBpedia ontology, given its broad semantic coverage and easy integration with the DBpedia Knowledge Base. We select the semantic annotations, which were obtained through embedding matching instead of syntactic matching~\cite{hulsebos2023gittables}, as we aim to have diverse and sufficient coverage. We filter out semantic annotations with a cosine similarity than 0.75\footnote{The cosine similarity is calculated from the similarity between column name and semantic type, indicating a certain level of confidence of the annotation~\cite{hulsebos2023gittables}.}.

In total, we train \system Predictor on columns from approximately 15K tables from GitTables. We further select tables so that each table includes at least 1 column that matches one of the 10 semantic types in our initial type catalog. Columns not matching any of the types are regarded as samples of the background class (labeled with type \type{null}). We train the \system Predictor on this background class to represent out-of-distribution data (see Figure~\ref{fig:adaptation-scenarios}c), a common method for representing non-discriminative ``intra-category variance''~\cite{dhamija2018reducing}. We downsample columns labeled with \type{null} to a sample size of 250 columns to avoid overfitting on the background class, whereas other overrepresented types are equally capped.


\subsection{\system Predictor}

\begin{figure}
    \centering
    \includegraphics[width=\columnwidth]{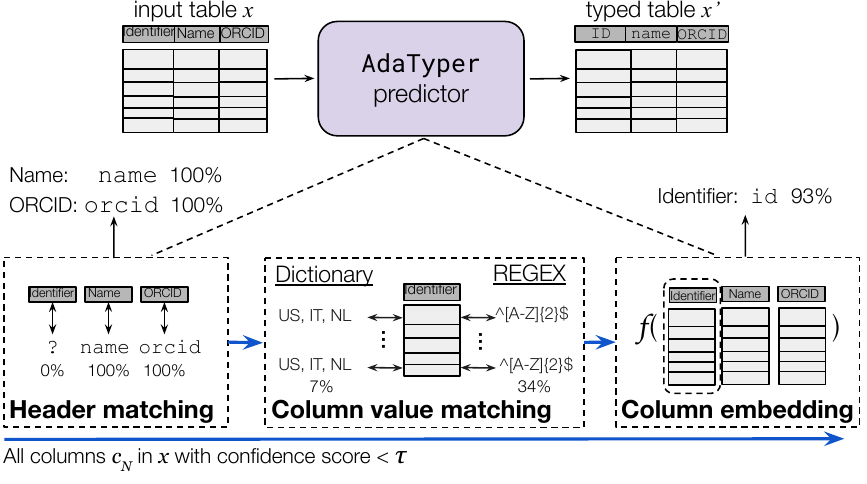}
    \vspace{-0.5cm}
    \caption{Overview of \system Predictor implementing a 3-step sequential pipeline to predict the semantic type of a column. Each estimator is executed only for columns that could not be predicted with sufficient confidence ($\tau$) by preceding estimators. Given a type catalog $\mathcal{T}$, the first estimator matches embeddings of the header with embeddings of the types in $\mathcal{T}$. The second estimator matches column values to preconfigured regular expressions and dictionaries. The last step uses an ML model trained on T5 column embeddings for the semantic type prediction task.}
    \label{fig:adatyper-predictor}
\end{figure}

The \system Predictor component implements a 3-step pipeline to predict the semantic types of columns, using signals from the header, column values, and embeddings of columns (Figure~\ref{fig:adatyper-predictor}).

Each estimator (step) in the pipeline is executed only if a preset confidence threshold $\tau$, which is specific to that estimator, is not met by the prior estimator. The first estimator only considers the header, whereas the second performs basic column value comparisons. The third and last estimator embeds columns with a pretrained language model. This sequential procedure reduces overhead by heavier estimators as the lightweight estimators, e.g. header matching, may already yield a prediction hence make full table scans redundant.

\paragraph{Header matching}
As the header of $x$ might accurately resemble the semantic type of a column, \system Predictor first matches each column name to the types in the type ontology using semantic matching. We leverage the light-weight Universal Sentence Encoder (USE)~\cite{cer2018universal}, which robustly embeds column names composed of multiple words. The confidence corresponds to the cosine similarity between the USE embedding of the column name and the embeddings of the types in type catalog $\mathcal{T}$.

\paragraph{Column value matching}
If the confidence of the header matching step does not surpass a preset threshold for a subset of columns, the next estimator is triggered to label them. This estimator leverages column value matching, to match a sample of column values to semantic types in $\mathcal{T}$. The matchers consist of 1) a set of regular expressions for structured types\footnote{The regular expressions are at: \url{https://github.com/madelonhulsebos/AdaTyper}}, 2) dictionaries of values commonly found in columns of each semantic type (populated from GitTables). The percentage of column values matched for a given type, is returned as the confidence for that type. The type with the highest confidence scores is returned as predictions for a column, but only returned as final type prediction if the confidence is higher than $\tau_{\text{regex}}$ or $\tau_{\text{dict}}$, respectively.

\paragraph{TypeT5}
If the confidence of prior estimators doesn't meet the preset threshold, the pipeline embeds $x$ using a learned type detection approach based on column embeddings extracted with the general-purpose language model T5~\cite{raffel2020exploring}. We leverage the implementation from the Observatory library~\cite{cong2023observatory} to serialize and truncate table columns, and aggregate token-embeddings to column-level embeddings. Alternatives for column-level embeddings, such as TaBERT~\cite{yin2020tabert} and Sherlock~\cite{hulsebos2019sherlock}, were considered but whereas TaBERT is found to not generalize well to out-of-distribution data~\cite{kayali2023chorus}, the representation of Sherlock is composed of multiple feature sets that require a training procedure, making it unsuitable for our adaptation component (Section~\ref{sec:adatyper-adaptation}).

We train a basic tree-based machine learning model---a random forest classifier from scikit-learn---on the T5 column embeddings of the annotated training data. For medium-sized type catalogs $\mathcal{T}$, tree-based classifiers yield sufficient predictive performance on rich vector representations (T5), while being more robust to smaller sample sizes and more efficient to train, compared to adding neural layers on top of T5. These criteria are important because this estimator is retrained in the adaptation component (Section~\ref{sec:adatyper-adaptation}). Higher capacity models could be considered for large type catalogs and in non-adaptive type systems.


\begin{figure}
    \centering
    \includegraphics[width=0.7\columnwidth]{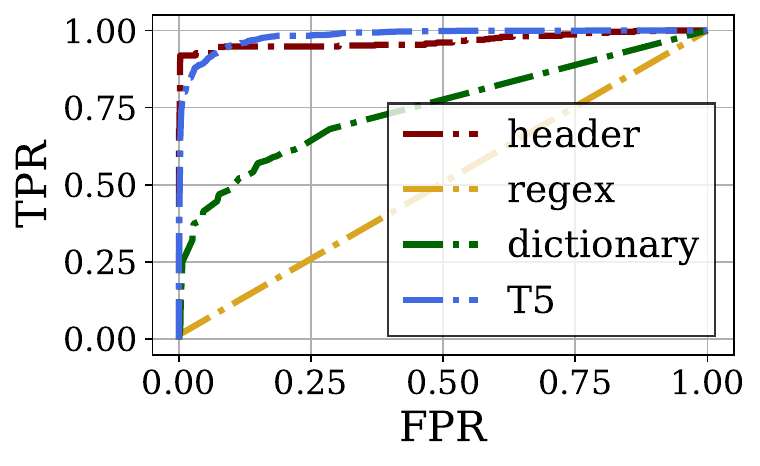}
    \vspace{-0.2cm}
    \caption{Evaluation of the trade-off between TPR and FPR for the header- and column value matchers (regular expressions and dictionary lookup) and TypeT5 estimators for different values of threshold $\tau$. We set $\tau_{\text{header}}$ to 0.75, $\tau_{\text{regex}}$ to 0.20, $\tau_{\text{dictionary}}$ to 0.27, and $\tau_{\text{TypeT5}}$ to 0.18.}
    \label{fig:tau-evaluation}
\end{figure}

\paragraph{Final type predictions}
Figure~\ref{fig:tau-evaluation} illustrates the trade-off between true positive rate (TPR) and false positive rate (FPR) per estimator. We set the confidence threshold $\tau$ for each estimator separately based on predictions on a hold-out set, such that the true positive rate is maximized while the false positive rate is close to 3\%. We consider this an acceptable FPR in deployment settings, as precision in user-facing systems is key to keeping users engaged and avoiding downstream issues. The values for $\tau$ may be set differently depending on system requirements. These thresholds are used to filter predictions with confidence scores below $\tau$, reflecting predictions with lower confidence than desired. The final prediction for each column in $x$ is the set of the types for which \system Predictor yields predictions with sufficient confidence. The \system Predictor does not provide a semantic type, i.e. predicts \texttt{null}, for columns if no estimator predicts a type with sufficient confidence.

As expected and shown in Figure~\ref{fig:tau-evaluation}, TypeT5 and the header-matching approach provide the highest performance in isolation. We set $\tau$ for each estimator such that it corresponds to an FPR close to 0.03. Table~\ref{tab:estimator-thres-perf} presents the thresholds ($\tau$) per estimator along with the associated FPR and TPR.



\begin{table}[]
    \centering
    \caption{True positive rate (TPR) and false positive rate (FPR) per estimator, with fixed $\tau$, showing the progression of TPR and FPR from resource-light to heavier estimators throughout \system Predictor.}
    \vspace{-0.2cm}
    \begin{tabular}{l c c c }
        \toprule
        \textbf{Estimator} & \textbf{threshold ($\tau$)}  & \textbf{FPR} & \textbf{TPR} \\
        \midrule
        Header matching\tablefootnote{The ground-truth labels of columns in GitTables are correlated with the column names due to the automated annotation process~\cite{hulsebos2023gittables}. The performance of this estimator is therefore positively biased; hence we set $\tau_{\text{header}}$ to 0.75 instead.} & 0.61 (0.75) & 0.03 & 0.92 \\
        Regex matching & 0.20 & 0.0001 & 0.01 \\
        Dictionary matching & 0.35 & 0.03 & 0.37 \\
        TypeT5 & 0.18 & 0.03 & 0.88 \\
        \bottomrule
    \end{tabular}
    \label{tab:estimator-thres-perf}
\end{table}

\begin{figure*}[h!]
    \centering
    \includegraphics[width=0.255\textwidth,valign=t]{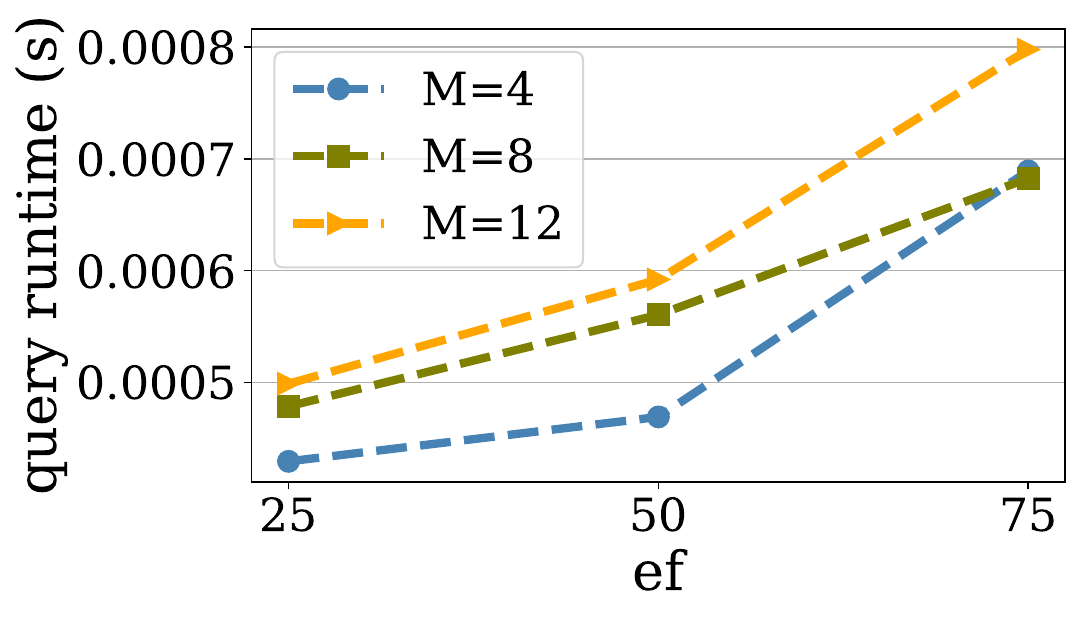}
    \includegraphics[width=0.24\textwidth,valign=t]{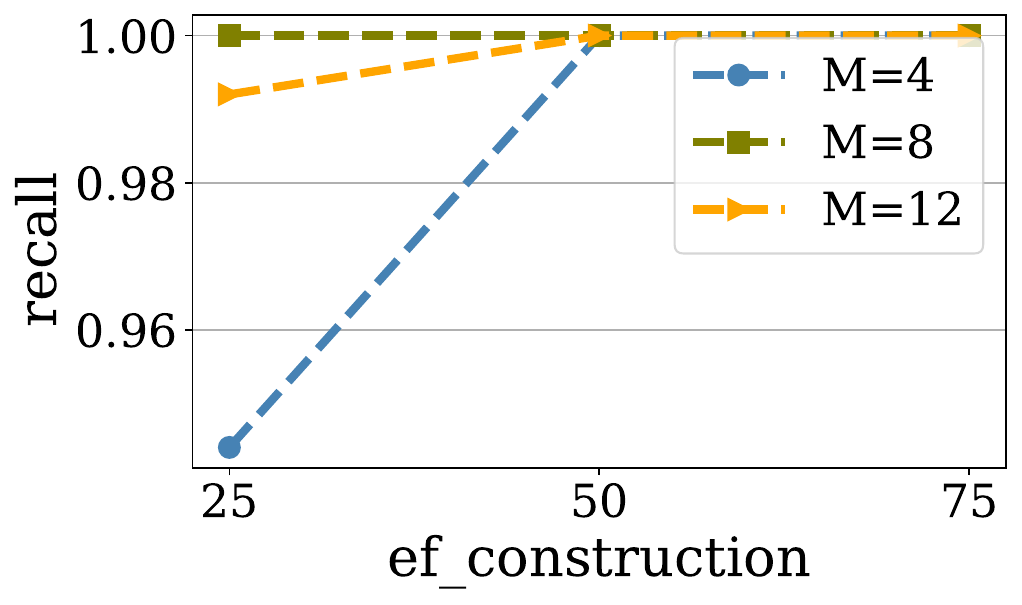}
    \includegraphics[width=0.242\textwidth,valign=t]{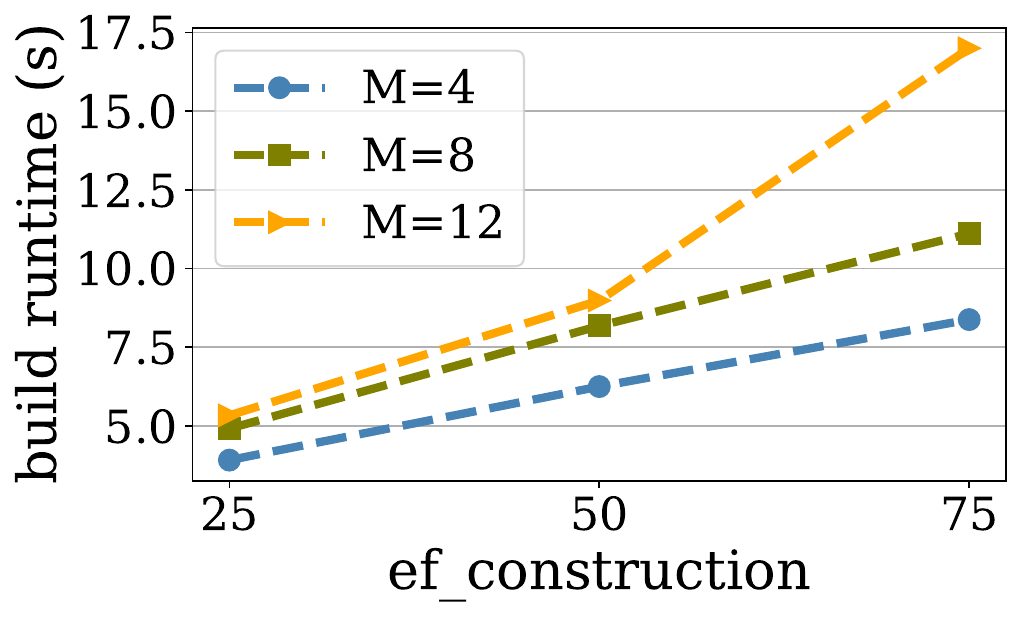}
    \includegraphics[width=0.242\textwidth,valign=t]{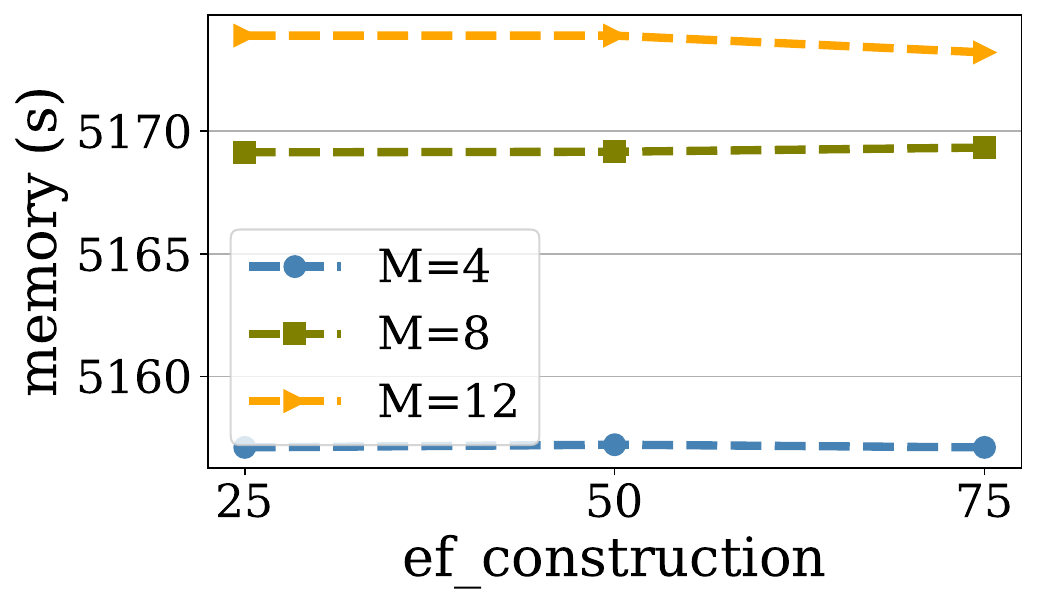}
    \vspace{-0.4cm}
    \caption{Evaluation of the HNSW index with T5 column embeddings in terms of retrieval recall, retrieval runtime, index memory usage, and construction time (which is a one-time procedure) over various 
    HNSW hyperparameter settings (fixing \texttt{ef}=50 or \texttt{ef\_construction}=50). Results are averaged over 5 runs. We set \texttt{M} to 8, and $\text{ef\_construction}$ and $\text{ef}$ to 50 as these settings yield high recall, efficient retrieval, index construction, and lowest memory usage (5.4GB).}
    \label{fig:hnsw-evaluation}
\end{figure*}

\subsection{\system Adapter}\label{sec:adatyper-adaptation}
To stay useful, systems incorporating learned models should adapt to the user's specific semantic context, without imposing time-consuming interaction on the user. Hand-labeling vast amounts of data to retrain machine learning models, especially deep learning models, is infeasible given budget constraints~\cite{stonebraker2018dataintegration}. Moreover, providing inputs for value matching, for example, through regular expressions or value dictionaries, requires expert knowledge of regular expressions or potentially occurring values. We therefore adopt a lean and scalable approach and make \system adapt by example using weak-supervision.

\paragraph{Adapting column value matchers}
For the column value estimator, we use two approaches to adaptation. First, the most common values found in the example column are added to the dictionary estimator for the new semantic type. This adaptation is executed automatically. Second, the user may provide a regular expression which is then used as the regular expression for the new or existing type. It is therefore added to \system its set of regular expressions.

\paragraph{Adapting TypeT5}
As shown in Figure~\ref{fig:adaptation-adatyper}, the adaptation cycle of \system adopts weak-supervision for generating training data with a retrieval approach requiring minimal user input. That is, we consider the column with the corrected semantic type (either a new unseen type, or an existing type from the type catalog $\mathcal{T}$) as an ``example'' of what columns of the given new semantic type are like. We use an index prefilled with columns from GitTables to retrieve columns similar to the example column. 

We use the T5 embedding representations of the columns as used for training \system Predictor, to extract an embedding of the example column\footnote{In an early prototype~\cite{hulsebos2022making}, we devised labeling functions (LFs) to represent the example column and training table columns. However, we found that without labeled data or prior knowledge of the semantic types that the system adapts to, the LFs cannot be calibrated to minimize noise in the training data, which affected prediction performance.}. The T5-embedding of the example column is the input for a retrieval procedure, which retrieves $k$ approximate nearest neighbors (ANN) from a Hierarchical Navigable Small World (HNSW) index~\cite{malkov2018efficient} built with T5 embeddings of columns from GitTables. The HNSW index is an efficient and accurate index for retrieving similar high-dimensional vectors given a query vector (the example column in the case of \system). We use cosine similarity as the metric for building the index and retrieving similar columns from it.

We evaluate several hyperparameter settings for the HNSW index as shown in Figure~\ref{fig:hnsw-evaluation}. Hyperparameter $\texttt{M}$ represents the number of neighbors that each ``node'' in the graph is connected to (resembling the degree of the node). Therefore, a higher $\texttt{M}$ yields higher recall, at the cost of memory usage. Parameter $\texttt{ef}$ specifies the number of neighbors considered as candidates for nearest neighbor lookup during retrieval, and influences the recall as well as query runtime as shown in Figure~\ref{fig:hnsw-evaluation}. With $\texttt{ef}$ set to a sufficiently high value, we find that the quality of the index $\texttt{ef\_construction}$ (the number of neighbors considered to prune the index at construction time) yields high recall at 50. We prioritize query runtime and recall, and set the hyperparameters $\texttt{M}$ to 8, $\texttt{ef}$ to 50, as this configuration yields high recall at minimum cost at retrieval (query) runtime.

The $k$ retrieved column embeddings are then labeled with the new semantic type and added to the training data, along with the example column. Finally, TypeT5 is retrained on the training data that now includes the generated weakly-supervised training data for the new type. The high-level procedure for adapting the \system Predictor towards new types and shifted data distributions (scenarios (a) and (c) in Fig.~\ref{fig:adaptation-scenarios}) using weak-supervision is summarized in Algorithm~\ref{alg:adaptation} and visualized in Figure~\ref{fig:adatyper-overview}.

\vspace{-0.1cm}
\begin{algorithm}
    \SetAlgoLined
    \DontPrintSemicolon
    \SetKwProg{Fn}{Function}{:}{}
    \SetKwFunction{FAdapt}{Adapt}
    \SetKwFunction{FIndex}{BuildIndex}
    \KwData{table $x$, example column $c$, (new) type $t$, type catalog $\mathcal{T}_i$, HNSW index $\mathcal{H}$, number of columns to retrieve $k$, \system predictor $P$}
    \KwResult{Enhanced type catalog $\mathcal{T}_{i+1}$, retrained $P_{i+1}$, enhanced train set $\mathcal{X}_{i+1}$}

    \Fn{\FAdapt{x, t, $\mathcal{T}$, $\mathcal{H}$, $P_i$}}{
    
        column embedding $c_e = \texttt{TypeT5}(x)[c]$ \;
        $\mathcal{C}_e$ = query($\mathcal{H}$, k) \;
        $\mathcal{X}_{i+1} = \mathcal{X}_{i} \cup \mathcal{C}_e$ \;
        $\mathcal{T}_{i+1} = \mathcal{T}_i \cup t$ \;
        $P_{i+1} = \text{retrain} \ P(\mathcal{X}_{i+1})$ \;
        
        \Return{$\mathcal{T}_{i+1}$, $\mathcal{X}_{i+1}$, $P_{i+1}$} \;
    }
    \caption{Adaptation of \system type predictor and the training dataset through weak-supervision.}
    \label{alg:adaptation}
\end{algorithm}

%% file: 06_evaluation.tex
\section{Experiments}\label{sec:results}

In this section, we analyze 1) the general performance and effectiveness of the \system Predictor and 2) the adaptability of \system Adapter towards new types and shifting data distributions on the hand-labeled out-of-distribution dataset.

\subsection{General performance evaluation}

\paragraph{Dataset}
A basic condition for a semantic type detection system is its initial performance on the semantic types and data distribution that the system is trained on. This evaluation is done with unseen tables from a hold-out test set from GitTables~\cite{hulsebos2023gittables}. We train the \system Predictor on 1.3K columns corresponding to 10 semantic types with general relevance, e.g. geographic types, business types, and types indicating Personal Identifiable Information (PII).

\paragraph{Experimental setup}
We evaluate the prediction performance with standard multi-class metrics precision, recall, and F1-score. To illustrate the performance of the most commonly deployed type prediction methods in industry applications, pattern matching, and dictionary lookup, we evaluate the estimators in isolation as well. Finally, we measure the percentage of predictions contributed by each estimator to illustrate their influence on the overall pipeline. The estimators are configured with confidence thresholds $\tau$ as in Table~\ref{tab:estimator-thres-perf}.

\begin{figure}
    \centering
    \begin{subfigure}[t]{0.48\columnwidth}
    \includegraphics[width=\linewidth]{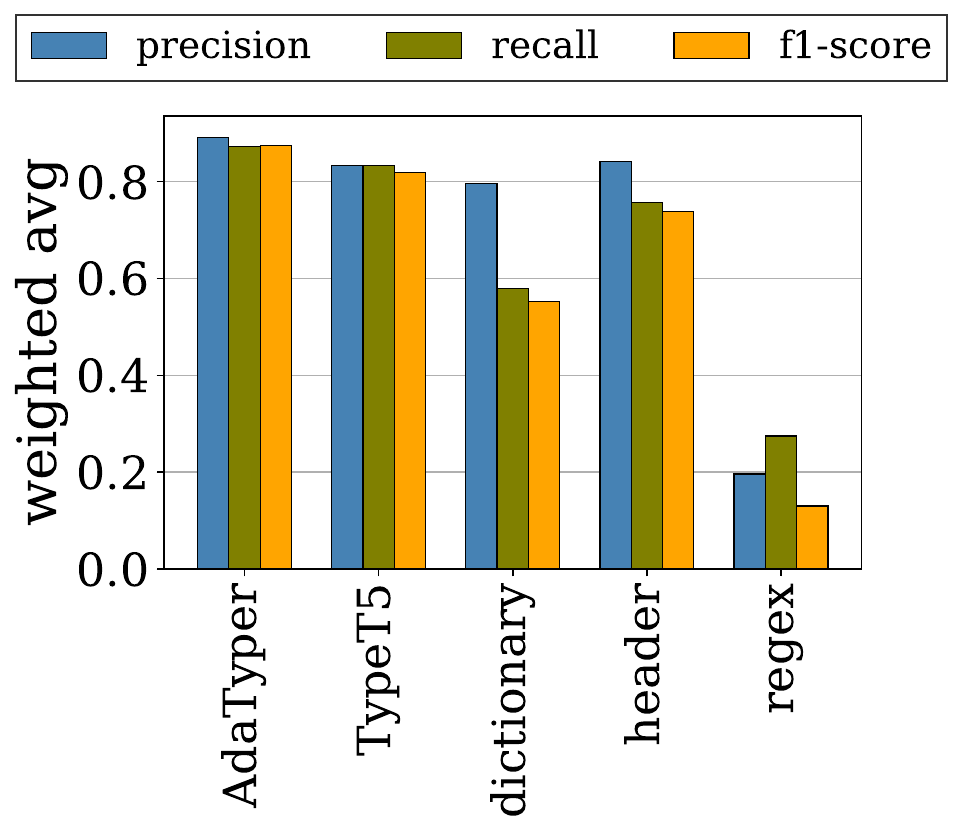}
    \vspace{-0.48cm}
    \caption{Performance of each estimator in isolation and the end-to-end pipeline of \system Predictor. While the value matching estimators yield high precision but lower recall on average (weighted), TypeT5 contributes the best performance in isolation, headed by the end-to-end \system Predictor.}
    \end{subfigure}\hfill
    \begin{subfigure}[t]{0.48\columnwidth}
    \includegraphics[width=.98\columnwidth]{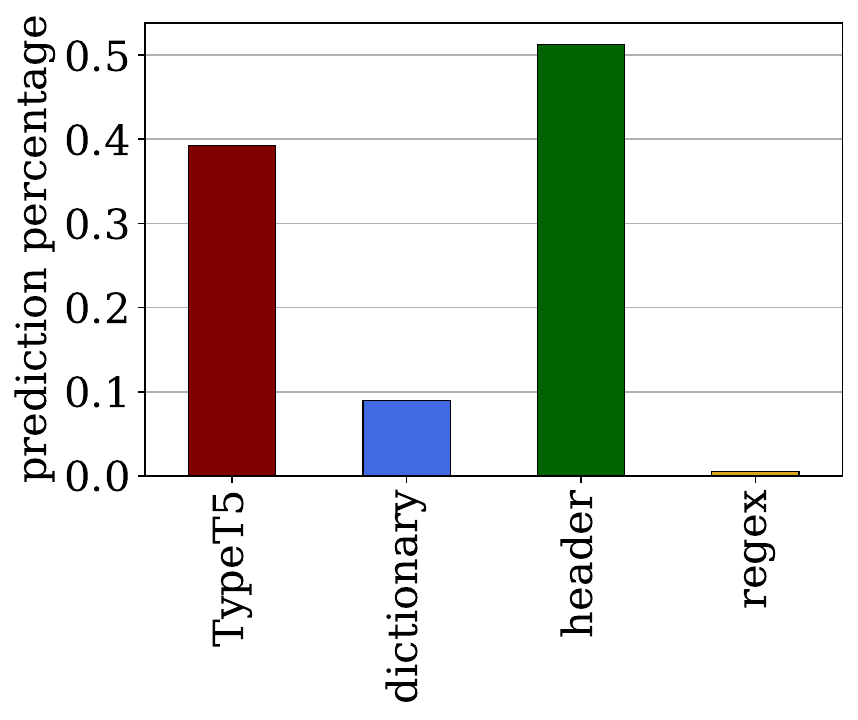}
    \vspace{-0.1cm}
    \caption{Percentage of estimator predictions generated by each estimator, reflecting its ``influence'' in the end-to-end pipeline of \system Predictor. After the biased header-based estimator, the ML-estimator contributes the most, while the dictionary estimator also has a significant share in the final prediction.}
    \end{subfigure}
    \vspace{-0.3cm}
    \caption{Evaluation of \system Predictor on an in-distribution unseen subset of GitTables. The end-to-end pipeline metrics correspond to the configuration of $\tau$ as in Table~\ref{tab:estimator-thres-perf}.}
    \label{fig:estimator-predictor-evaluation}
    \vspace{-0.3cm}
\end{figure}

\paragraph{Results} As Figure~\ref{fig:estimator-predictor-evaluation}(a) shows, the value matching estimators yield high precision but relatively lower recall. TypeT5 yields the best performance in isolation across the three metrics, with all scores around 0.8. Given the correspondence of the annotations and the column names, the header-based estimator yields good performance, too, albeit not guaranteed to generalize beyond GitTables. The hybrid prediction pipeline, \system Predictor, effectively exploits the strengths of each estimator and consistently yields the best prediction performance. In Figure~\ref{fig:estimator-predictor-evaluation}(b), we demonstrate the contribution of each estimator towards the final prediction in the \system Predictor, in terms of the percentage final predictions generated by the respective estimator. Similarly, the header-estimator and TypeT5 are key contributors, while the dictionary estimator significantly contributes a significant portion as well.

\subsection{Performance \system Adapter}

\begin{figure}
    \centering
    \shadowbox{\includegraphics[width=0.9\columnwidth]{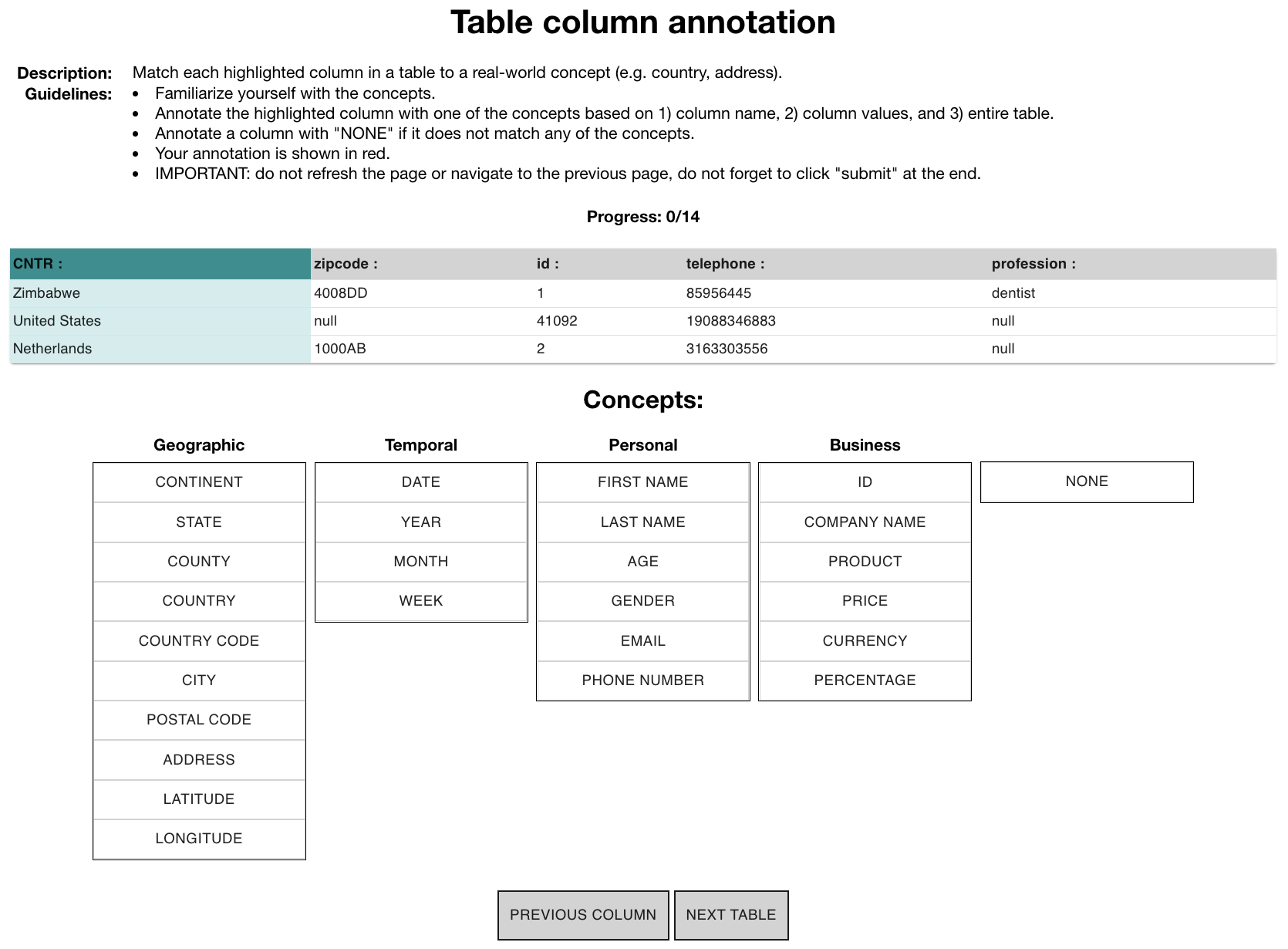}}
    \vspace{-0.1cm}
    \caption{The TableAnnotator user-interface for obtaining semantic column type annotations for tables from the CTU Prague Relational Learning Repository. After preprocessing, we obtain 2,665 hand-labeled columns distributed over 26 non-\type{null} semantic type annotations as in Figure \ref{fig:annotation-distribution}.}
    \label{fig:table-annotator}
    \vspace{-0.1cm}
\end{figure}

\paragraph{Dataset} Transferring towards real-world database tables is key for systems deployed in enterprise data analysis tools. In the absence of a human-annotated corpus of real-world database tables, we collect and annotate 1K tables with, on average, 8 columns from the CTU Prague Relational Learning Repository~\cite{motl2015ctu}. The CTU PRLR corpus is not included in GitTables, hence provides an appropriate dataset for evaluating \system Adapter on its effectiveness on out-of-distribution data. Given the scale of the CTU dataset, we use crowd-sourcing platform Mechanical Turk (MTurk) to obtain semantic column type annotations for each of the columns in these tables. We developed the TableAnnotator web app\footnote{The code can be found at: \url{https://github.com/madelonhulsebos/TableAnnotator}.} (see Figure~\ref{fig:table-annotator}) to facilitate the column type annotation task. The TableAnnotator can easily be reused to annotate any set of tables with any set of types, or extended to other table annotation tasks.

We split the CTU PRLR dataset into sets of 13 tables, add a practice table for training, and a honeypot table with known ground-truth annotations to evaluate annotation quality. Every Worker, therefore, annotates 15 tables in total. Each set is annotated by 5 different Workers to get 5 annotations per table column for post-hoc aggregation. We evaluated various experimental designs for the annotation task (e.g. Workers with an average HIT approval rate of 99\%) that resulted in different quality levels as measured by the performance on the honeypot table (see Table~\ref{tab:annotation-setup}). We adopted MTurk's automatically inferred Master qualification\footnote{A qualification given by MTurk to Workers with the highest annotation performance.}.

\begin{table}[]
    \centering
    \caption{Annotation quality across different experimental designs (average HIT approval rate, number completed HITs, Master qualification). We select the Master qualification configuration for its highest precision and f1-score.}\label{tab:annotation-setup}
    \vspace{-0.2cm}
    \begin{tabular}{l c c c c}
    \toprule
	\textbf{Configuration} & \textbf{precision} & \textbf{recall} & \textbf{f1} \\
	\midrule
    Master qualification & \textbf{0.93} & 0.78 & \textbf{0.82} \\ 
    approval > 99\% \#completed > 100 &	0.81 & \textbf{0.82} & 0.80 \\ 
    approval > 98\% \#completed > 1000 &	0.89 & 0.57 & 0.65 \\ 
    \bottomrule
    \end{tabular}
    \vspace{-0.2cm}
\end{table}

We further control  annotation quality by (post-hoc) filtering out annotations from Workers that do not have the null type in their top-$k$ most common annotation labels, indicating random clicking. We aggregate the resulting set of annotations $(a_1,...a_n)$ into the final type annotation $t$ for a given column by taking the most frequent type, if its frequency is higher than the minimum vote ($\text{min\_vote}$), as displayed in Equation~\ref{eq:annotation-aggregation}.
\begin{equation}\label{eq:annotation-aggregation}\small{
t = \begin{cases}
    \text{majority}(a_1...a_n) & |\text{majority}(a_1...a_n) \in a_1...a_n| \geq \text{min\_vote} \\
    \text{null} & \text{otherwise}
\end{cases}}
\end{equation}

\begin{figure}
    \centering
    \includegraphics[width=0.9\columnwidth]{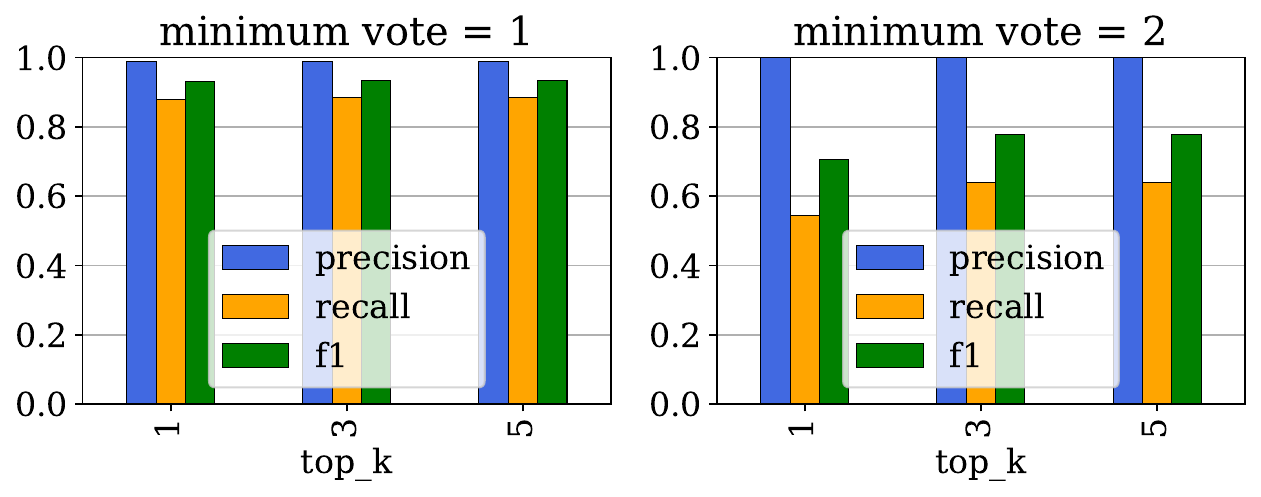}
    \vspace{-0.3cm}
    \caption{The post-hoc annotation quality as aggregated with various settings for the top-$k$ most frequent type and $\text{min\_vote}$ for aggregation as expressed in equation \ref{eq:annotation-aggregation}. We filter out the \type{null} annotations from these aggregate scores. We set $\text{min\_vote}$ 1 and $k$ to 3, respectively, as all configurations yield similar precision while recall is much higher compared to $\text{min\_vote}$=2.}
    \label{fig:annotation-aggregation}
    \vspace{-0.1cm}
\end{figure}

\begin{figure}
    \centering
    \hspace{-0.1cm}
    \includegraphics[width=0.95\columnwidth]{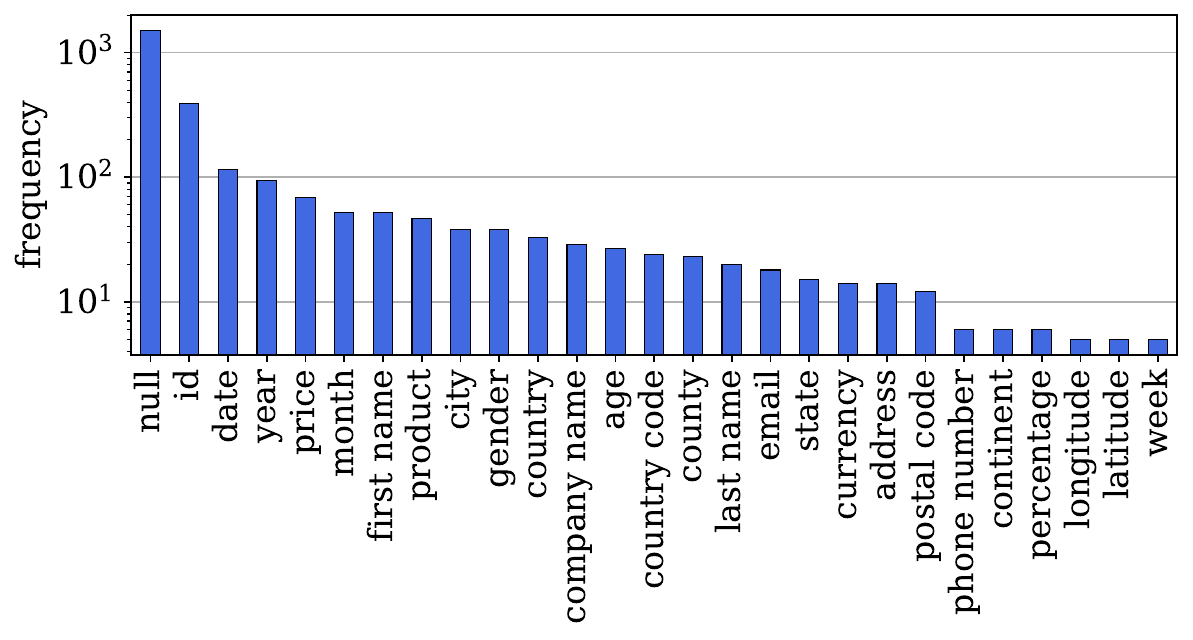}
    \vspace{-0.3cm}
    \caption{The distribution of semantic types with hand-labeled annotations in the processed evaluation CTU PRLR dataset. As typical in databases tables, the \texttt{id} type is the most common type.}
    \label{fig:annotation-distribution}
\end{figure}

We evaluate the annotation quality over different values for $k$ and $\text{min\_vote}$ as presented in Figure~\ref{fig:annotation-aggregation}. We adopt $k=3$ and $\text{min\_vote}=1$ as all configurations yield similar high precision while the recall, and therefore the number of non-\type{null} annotations, is significantly higher when $\text{min\_vote}$ is set to 1 compared to 2. After aggregation with the final configuration and preprocessing to extract column-level embeddings we obtain 2,665 hand-labeled table columns associated with the annotations distributed as shown in Figure~\ref{fig:annotation-distribution}.

\paragraph{Experimental setup}
We evaluate how well \system adapts to new types given only a handful of example columns of the new type. These examples can be obtained from users through any user-interface with a tabular component, no need for defining regular expressions, approving them, or providing example values. In this experiment, the tables for the ``example columns'' of the new types are sourced from the hand-labeled CTU PRLR dataset to reflect the target distribution. We present a fine-grained analysis of the adaptation performance of \system Adapter across multiple types and metrics. Finally, we compare the performance of \system Adapter with the baseline performance obtained with two adaptation methods commonly used in practice.

We evaluate the predictive performance across 5 adaptation cycles (adapting to one example column per cycle) for 4 different new types: \type{first name}, \type{city}, \type{gender}, and \type{postal code}. We remove the example columns from the CTU evaluation set to prevent training data leakage. The types were selected such that we have at least 10 samples in the CTU dataset to have enough examples for adaptation as well as evaluation. Three types are new types that \system Predictor was not trained on (out-of-distribution adaptation, Figure~\ref{fig:adaptation-scenarios}(a)). Two out of four types (\type{city} and \type{gender}) already exist in the type catalog on which \system Predictor was trained, illustrating adaptation towards existing types with a shifting data distribution (covariate shift, Figure~\ref{fig:adaptation-scenarios}(a)). To enable a level of fair comparison with the baselines, we ensured that for two types (\type{gender}, and \type{postal code}) we could find at least one regular expression\footnote{Many semantic types, e.g. \type{company}, are less suitable for regex-matching as they lack distinctive syntactic patterns. This setup therefore slightly favors the regex baseline.}.

\begin{figure}%
    \centering
    \vspace{-0.1cm}
    \subfloat[\centering \system adaptation performance by different metrics\label{fig:adaptation-adatyper}]{{\includegraphics[width=0.46\columnwidth,valign=t]{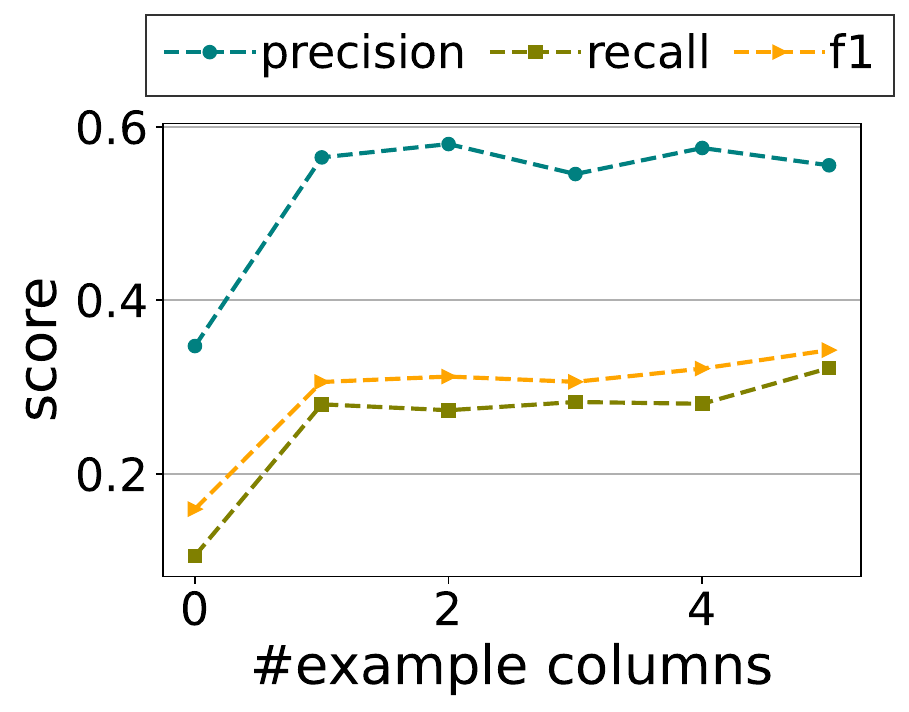} }}%
    \subfloat[\centering \system adaptation comparison with baselines\label{fig:adaptation-baselines}]{{\includegraphics[width=0.5\columnwidth,valign=t]{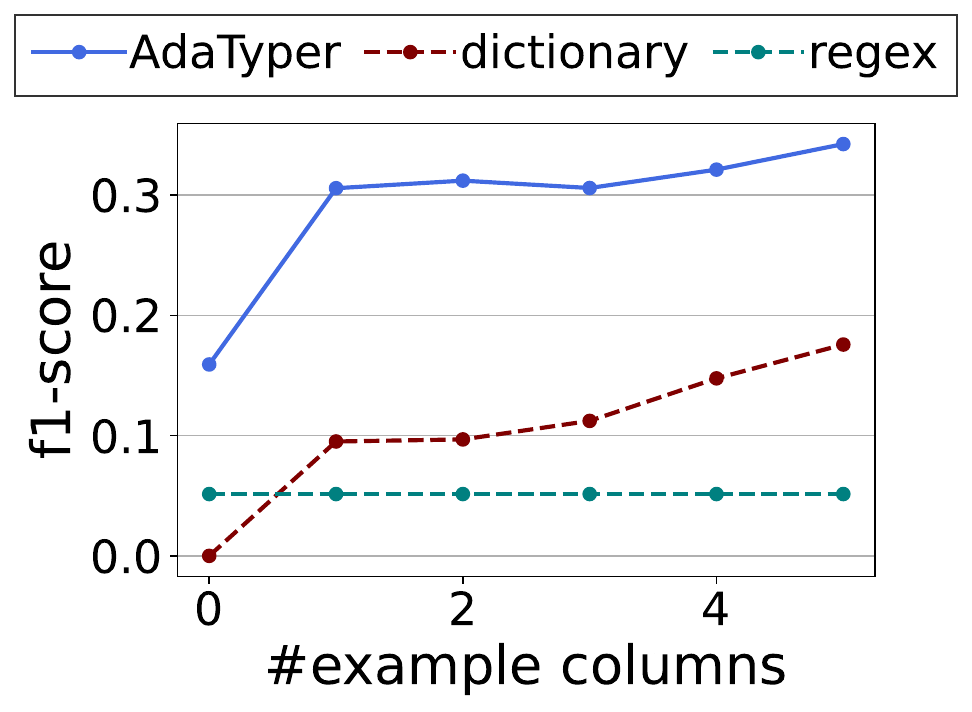} }}%
    \vspace{-0.2cm}
    \caption{Adaptation performance evaluation, showing how (a) \system Adapter improves over the number of examples seen due to an increase in recall, and (b) \system Adapter iteratively improves in f1-score in comparison with the regex- and dictionary baselines which plateau at an f1-score of 0.1-0.15.}%
    \vspace{-0.3cm}
\end{figure}

\paragraph{Results}
Figure ~\ref{fig:adaptation-adatyper} shows that, on average, \system is able to reasonably detect the new types after seeing already one example column of a new type. As new examples are provided and new weakly-supervised training data is generated, the recall hence F1-score increases on every iteration. This implies that more columns of the new type are recognized by \system. After only labeling 5 examples, the ability of \system to identify columns of the new types goes up from 0 to 0.3, on average, as indicated by the rising recall curve. Due to the weakly supervised generated training data, the precision only lightly improves or slightly degrades depending on the representatitiveness of the example columns, steadily approaching an average precision of 0.6 after 5 examples. We observe a high variance in the performance across the types.

We compare \system Adapter to adaptation methods commonly used in practice~\cite{googledatastudio:semantic-type, trifacta:type-system}. As baselines, we consider a method based on dictionary- and regular expression matching for which we reuse the dictionary- and regex-based estimators as in \system Predictor. For the dictionary baseline, we populate a dictionary of $k$ most common values per type from the example columns, the dictionary increases in size as new examples are added to the system. For the regex baseline we retrieve regular expressions from the Web. 

\begin{figure}[]
    \centering
    \includegraphics[width=0.329\columnwidth,valign=t]{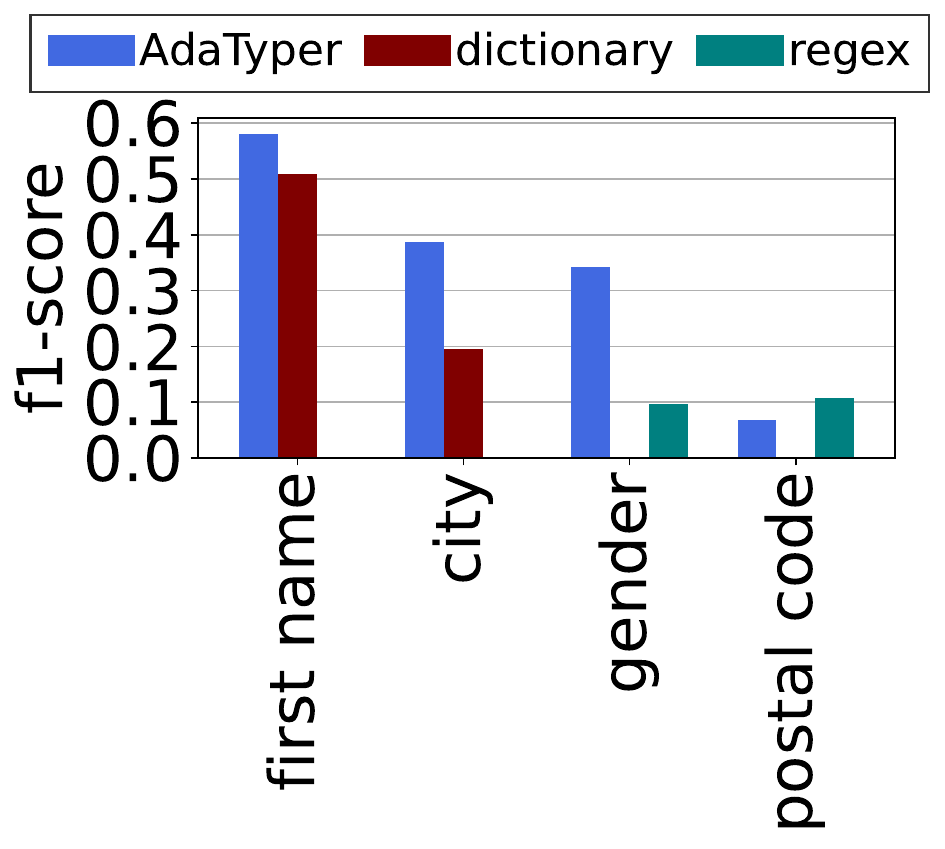}
    \includegraphics[width=0.329\columnwidth,valign=t]{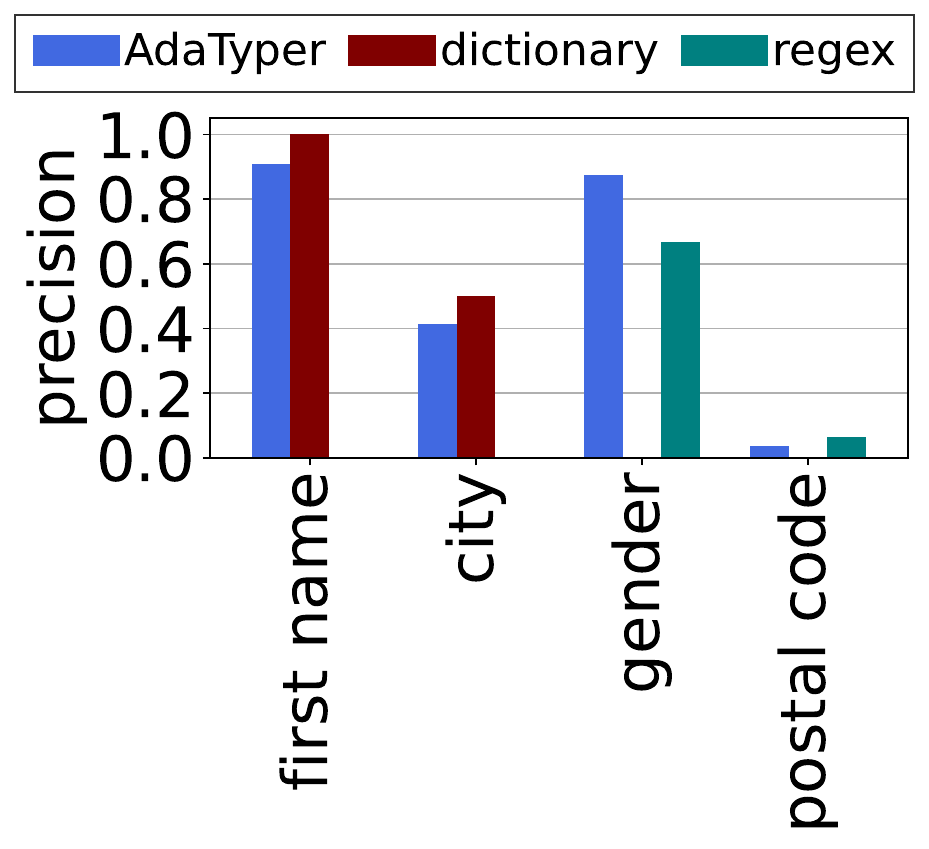}
    \includegraphics[width=0.329\columnwidth,valign=t]{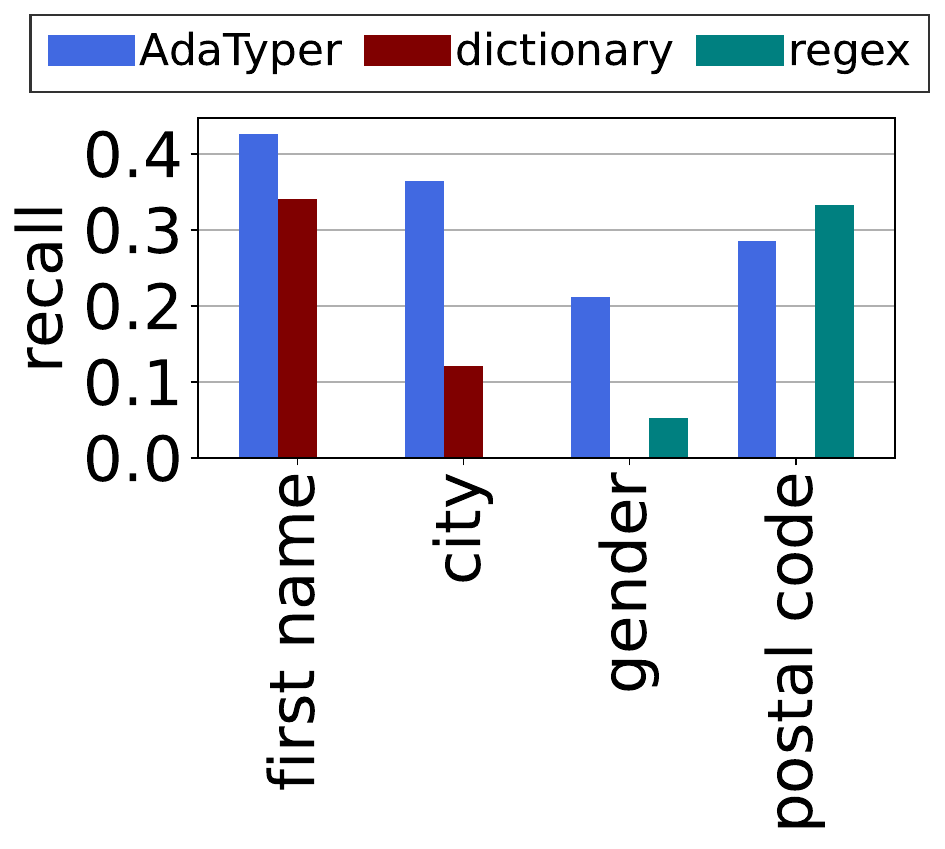}
    \vspace{-0.1cm}
    \caption{Per-type performances for \system Adapter after 5 examples, with high variance across different types and metrics. Notably, adapting to textual types such as \texttt{first name} is effective, resulting in an f1-score close to 0.6. Due to the difficulty of numeric data it is harder to adapt to semantic types such as \texttt{postal code} with precision, for which regular expressions are slightly better.}
    \label{fig:adaptation-zoomin}
\end{figure}

For the baseline adaptation methods, we find that the dictionary-based baseline only slightly improves upon being provided with more example columns approaching an f1-score close to 0.2 after 5 examples (Figure~\ref{fig:adaptation-baselines}). The regex-based approach, which is value-independent, yields a constant f1-score of 0.07 and is outperformed by both approaches. Moreover, due to the diversity of the semantic types, we observe a high variance across semantic types for all adaptation methods as presented in Figure~\ref{fig:adaptation-zoomin}. While \system Adapter achieves better precision for textual semantic types, between 0.4 and 0.9, its precision on numeric types such as \texttt{postal code} is much lower, affecting the average performance metrics as shown in Figure~\ref{fig:adaptation-adatyper}.

Inspection of the underlying performance measures at the end of the adaptation in Table~\ref{tab:adaptation-evaluation} reveals that both baselines achieve good precision but low coverage as the recall after 5 examples is on average 0.32. Altogether, \system Adapter improves the prediction performance over all types, ranging from f1-score deltas of +0.001 to +0.580. We conclude that adapting to new semantic types through weakly supervised retrieval of similar table embeddings is the most effective method for lightweight adaptation.

\begin{table}[]
    \centering
    \vspace{-0.3cm}
    \caption{Per-type f1-scores after adapting through 5 examples (i=5) with dictionary- and regex-based baselines and \system with a comparison to initial detection performance (i=0), if any.}\label{tab:adaptation-evaluation}
    \vspace{-0.1cm}
    \begin{tabular}{l c c c c l}
    \toprule
	\textbf{Type} & \textbf{regex} & \textbf{dictionary} & \multicolumn{3}{c}{\textbf{AdaTyper}} \\
                           &           &      &  i=0 & i=5 & $\Delta$F1\\
	\midrule
    \type{first name}  & 0  & 0.508 & - & \textbf{0.580} & +0.580 \\
    \type{postal code} & \textbf{0.108}  &  0  & - & 0.068 & +0.068 \\
    \type{city}         & 0.195 &  0  & 0.296 & \textbf{0.387} & +0.091\\
    \type{gender}        & 0  & 0.098 & 0.340 & \textbf{0.341} & +0.001 \\
    \bottomrule
    \end{tabular}
    \vspace{-0.3cm}
\end{table}

%% file: 07_related_work.tex
\section{Related Work\label{sec:related}}

\system builds on prior approaches to semantic column type detection that we discuss below. While its ensemble predictor partly overlaps with these approaches, \system is unique because it provides a semantic type prediction pipeline to predict, adapt, and retrain based on lightweight end-user feedback.


\paragraph{Matching based}
A common method for semantic type detection is to syntactically match values in a given column with regular expression patterns or reference (dictionary) exemplars that correspond to predefined types. Commercial data analysis and preparation systems such as  Microsoft Power BI~\cite{powerbi}, Google Data Studio~\cite{googledatastudio}, Trifacta~\cite{trifacta}, and Talend~\cite{talend} generally rely on this approach. Akin to existing systems, \system's column-value matcher also employs regular expressions and dictionary lookups.

\paragraph{Synthesized}
Another common approach is to autogenerate detection rules from examples. For instance, Yan and He~\cite{yan2018synthesizing} synthesize type detection logic from open-source GitHub repositories using a search keyword along with positive examples. Auto-Tag~\cite{he2021auto} synthesizes higher-level pattern-matching expressions to label table columns with predefined standard types. Beyond type detection, generating domain-specific language expressions to match examples is common in data preparation systems~\cite{raman2001potter,trifacta}.

\paragraph{Learned}
An increasingly more popular line of work uses machine learning, including graphical~\cite{goel2012exploiting, limaye2010annotating,takeoka2019meimei}, shallow-~\cite{li1994semantic} and deep neural networks~\cite{hulsebos2019sherlock, chen2019learning, takeoka2019meimei, zhang2020sato, wang2021tcn}, and pretrained~\cite{turl, iida2021tabbie, suhara2021annotating} models.  Prior research also uses hybrid~\cite{zhang2020sato, chen2019learning} and ensemble models~\cite{puranik2012specialist}. For example, Puranik~\cite{puranik2012specialist} combines the predictions of ``experts,'' including regular expressions, dictionaries, and machine learning models. \system’s predictor is an ensemble model combining value-matching and pretrained models. \system differs, however, from this earlier work by providing an end-to-end semantic type detection pipeline to support adaptation.

\paragraph{Adaptive} Techniques for adapting semantic type detectors implemented in data preparation tools are typically expert-driven. For example, Trifacta~\cite{trifacta:type-system} enables detecting custom types through user-specified regular expressions or value dictionaries, requiring expertise in data and pattern matching.  Auto-Tag~\cite{he2021auto} synthesizes pattern-matching expressions from values in table columns for user-specified types to extend the type predictor beyond predefined standard types. Similar to Auto-Tag, \system supports detecting custom types but goes beyond value-pattern matching methods, which typically over- or undergeneralize. Instead, \system represents columns with learned embeddings, providing more accurate and robust adaptive performance.

%% file: 08_conclusion.tex
\vspace{-0.2cm}
\section{Conclusion}\label{sec:conclusion}
\vspace{-0.1cm}
Learned table representation models show promising performance for detecting column types, entities, relations, and table topics. We believe the time has come to make these advances more accessible and explore avenues to make learned table representation models work in practice. In this work, we address one of the key challenges in practice, namely adaptation in downstream tasks such as semantic column type detection. Existing systems typically adapt through user-provided regular expressions or custom dictionaries, which rely on human expertise, over- or under-generalize, and do not scale well. 

We present \system that adapts to unseen types and shifted data distributions by leveraging learned table embeddings and weak-supervision. \system adapts using minimal input; it requires a system's user only to correct a type prediction with an existing or new type. This column is taken as an example that is used for retrieving similar columns, which are then added to the training data. The lightweight adaptation framework is effective and introduces many avenues for future research.

%% file: acknowledgements.tex
\begin{acks}
This project was supported by Sigma Computing.
\end{acks}